\RequirePackage[l2tabu, orthodox]{nag}
\documentclass[conference,a4paper]{IEEEtran}

\usepackage{amsmath}
\usepackage{amssymb}
\usepackage{wasysym}
\usepackage{graphicx}
\usepackage{subfigure}
\usepackage{color}
\usepackage{bm,cite}
\usepackage{tikz}
\usepackage{mathrsfs}
\usepackage{multirow}
\usepackage{array}
\usepackage{booktabs}
\usepackage{stfloats}
\usepackage{enumerate}
\usepackage{microtype}

\title{Enhanced Trellis Coded Multiple Access (ETCMA)}

\author{\IEEEauthorblockN{Alberto~G.~Perotti and Branislav~M.~Popovi\'c}\\
\IEEEauthorblockA{
Huawei Technologies Sweden AB \\
Skalholtsgatan 9-11, SE--164 94 Kista, Sweden \\
E-mails: \texttt{\small\{alberto.perotti, branislav.popovic\}@huawei.com}}}

\newcommand{\maxs}{\mathop{{\rm max}^*}}

\begin{document}

\maketitle

\begin{abstract}
We propose an enhanced version of trellis coded multiple
access (TCMA), an overloaded multiple access  scheme
that outperforms the original TCMA in terms of achieved
spectral efficiency.
Enhanced TCMA (ETCMA) performs simultaneous transmission of
multiple data streams intended for users experiencing
similar signal-to-noise ratios and can be employed both
in the \emph{uplink} and in the \emph{downlink} of wireless
systems, thus overcoming one of the main limitations of TCMA.
Thanks to a new receiver algorithm, ETCMA is capable of
delivering a significantly higher spectral efficiency.
We show that ETCMA approaches the capacity of the Additive
White Gaussian Noise channel for a wide range of
signal-to-noise ratios.
\end{abstract}

\section{Introduction}

Next generation wireless systems will have to face the demand
for higher aggregate data rates while being capable of providing
reliable communication to many simultaneous users and applications.
They will not necessarily be able to deliver higher \emph{per-user} data
rates, but higher \emph{aggregate} data rates to a growing number of users.
To this purpose, novel efficient Multiple Access (MA) techniques
are required.
Combined with high-performance coding and modulation schemes,
these techniques will be able to achieve higher aggregate and
per-user Spectral Efficiencies (SE) by means of a more
efficient use of the channel's physical resources.
Conventional MA techniques typically perform orthogonal multiplexing
of several coded and modulated data streams intended for
different users.
Each stream is transmitted using a different set of physical
time-frequency-space Resource Elements (REs), i.e. without
using REs for simultaneous transmissions of multiple streams.

\emph{Overloading} is a paradigm according to which, in a multiuser
transmission system, the number of users is greater than the
dimension of the signal space~\cite{bib:Kapur03TranIT}.
In the context of Code-Division Multiple Access (CDMA), overloading means using a number of
spreading sequences larger than the length of each sequence.
This concept has been recently extended beyond the context
of CDMA and new OverLoaded Multiple Access (OLMA) schemes
have been devised. In OLMA schemes, several data streams are
multiplexed and transmitted using the same REs in order to
provide increased data rates by achieving significantly higher
spectral efficiency than conventional MA schemes.

In our scenario, the optimization target is the maximization
of the aggregate Down-Link (DL) spectral efficiency (of one transmitter)
by simultaneous transmission to user equipment (UE) devices
experiencing similar physical channel qualities.
UEs that report to the DL transmitter the same channel quality
indicator value are grouped by the scheduler into the same
category, and then served by the same set of REs.
The corresponding OLMA methods preserve the same data rate,
the same transmitted energy per bit of each multiplexed stream,
and the same scheduler design as if each of the multiplexed
streams would have been transmitted alone on the observed
time-frequency-space resources.
It further means that the transmitted power per RE is increased
proportionally to the overloading factor, i.e. the number of
multiplexed streams.
OLMA schemes designed using this principle include, for
example, Low-Density Spread (LDS) MA
(LDSMA)~\cite{bib:Choi04ISSSTA, bib:BeekP09Globecom,
bib:Popov14WCNC}, Interleave-Division MA
(IDMA)~\cite{bib:PingL06} and Trellis-Coded MA
(TCMA)~\cite{bib:Aulin99ICC, bib:Brann02TCom}.

All the aforementioned schemes perform stream multiplexing
by superposition of coded and modulated signals by adopting
different solutions.
IDMA adopts a Bit-interleaved Coded Modulation
(BICM)~\cite{bib:BICM} approach with very low-rate codes.
LDSMA schemes use low-density signatures to enable detection
by low-complexity receivers.
The aggregate SE of these schemes approaches the Additive
White Gaussian Noise (AWGN) channel 
capacity in the low Signal-to-Noise Ratio (SNR) region, where the maximum SE is limited
to few bits/s/Hz. In the high-SNR region their SE 
diverges from the AWGN
channel capacity, thus requiring a
significant increase of transmitted power in order to reach high
aggregate spectral efficiency.

We fill this gap by proposing a new MA scheme based on TCMA
that, thanks to a carefully designed transmission scheme and to
an improved  receiver algorithm, exhibits a performance close to
the AWGN channel capacity for a wide range of SNR values and
features aggregate SE values as high as 7  bits/s/Hz.
Moreover, the new OLMA scheme can be used both in the \emph{downlink} and
in the \emph{uplink} of wireless systems.

This paper is organized as follows:
Sec.~\ref{sec:ETCMA} describes the ETCMA transmitter and receiver
schemes, Sec.~\ref{sec:results} presents the obtained results and
Sec.~\ref{sec:conclusions} draws the final conclusions.

\begin{figure*}[t]
\centering
\hfill
\subfigure[][Transmitter.]{
\begin{tikzpicture}[scale=.7]
\draw[rounded corners=5pt, line width=2.pt] (0.5,-.5) rectangle (8.125,6.5);
\node (tx) at (4.375,6.) {\bf\textsf{ETCMA Transmitter}};

\draw[->,line width=1pt] (.25,4.5) -- (1.,4.5) node [near start, above left=-2pt] {\small $u_0(l)$};
\draw[line width=1pt]  (1.,4.) rectangle (3.,5.);
\node at (2.0,4.5) {\small\textsf{${\rm TCM}_0$}};
\draw[->,line width=1pt] (3.,4.5) -- (3.5,4.5);
\draw[line width=1pt]  (3.5,4.) rectangle (4.75,5.);
\node at (4.125,4.5) {\small\textsf{$\rm \Pi_0$}};
\draw[->,line width=1pt] (4.75,4.5) -- (6.5,4.5) node [very near start, above right=-2pt] {\small $s_0(l)$};
\draw[line width=1pt] (6.75,4.5) circle (.25);
\node at (6.75,4.5) {$\times$};
\draw[->,line width=1pt] (6.75,3.75) -- +(0,.5) node [very near start, left=-1pt] {\small $c_0(l)$};

\draw[line width=.75pt, dotted] (2.,3.65) -- ++(.0,-.3);
\draw[line width=.75pt, dotted] (4.125,3.65) -- ++(.0,-.3);

\draw[->,line width=1pt] (.25,2.5) -- (1.,2.5) node [near start, above left=-2pt] {\small $u_k(l)$};
\draw[line width=1pt]  (1.,2.) rectangle (3.,3.);
\node at (2.0,2.5) {\small\textsf{${\rm TCM}_k$}};
\draw[->,line width=1pt] (3.,2.5) -- (3.5,2.5);
\draw[line width=1pt]  (3.5,2.) rectangle (4.75,3.);
\node at (4.125,2.5) {\small\textsf{${\rm \Pi}_k$}};
\draw[->,line width=1pt] (4.75,2.5) -- (6.,2.5) node [very near start, above right=-2pt] {\small $s_k(l)$};
\draw[line width=1pt] (6.25,2.5) circle (.25cm);
\node at (6.25,2.5) {$\times$};
\draw[->,line width=1pt] (6.25,1.75) -- +(0,.5)  node [very near start, left=-1pt] {\small $c_k(l)$};

\draw[line width=.75pt, dotted] (2.,1.65) -- ++(.0,-.3);
\draw[line width=.75pt, dotted] (4.125,1.65) -- ++(.0,-.3);

\draw[->,line width=1pt] (.25,.5) -- (1.,.5)  node [near start, above left=-2pt] {\small $u_{K-1}(l)$};
\draw[line width=1pt]  (1.,0.) rectangle (3.,1.);
\node at (2.0,.5) {\footnotesize ${\rm TCM}_{K-1}$};
\draw[->,line width=1pt] (3.,.5) -- (3.5,.5);
\draw[line width=1pt]  (3.5,0.) rectangle (4.75,1.);
\node at (4.125,0.5) {\small\textsf{${\rm \Pi}_{K-1}$}};
\draw[->,line width=1pt] (4.75,0.5) -- (6.5,0.5)  node [very near start, above right=-2pt] {\small $s_{K-1}(l)$};
\draw[line width=1pt] (6.75,0.5) circle (.25cm);
\node at (6.75,0.5) {$\times$};
\draw[->,line width=1pt] (6.75,-.25) -- +(0,.5)  node [very near start, left=-1pt] {\small $c_{K-1}(l)$};

\draw[line width=1pt] (7.375,2.5) circle (.35cm);
\node at (7.375,2.5) {$\Sigma$};
\draw[->,line width=1pt] (6.5,2.5) -- +(.5,0);
\draw[->,line width=1pt] (7.,4.5) -- ++(.375,0) -- +(0,-1.65);
\draw[->,line width=1pt] (7.,.5) -- ++(.375,0) -- +(0,1.65);
\node[anchor=north] at (9.,2.5) {$s(l)$};

\draw[line width=1pt] (7.75,2.5) -- ++(1.,0.) -- ++(0., 2) -- ++(-.25, 0.3) -- ++(0.25, -0.3) -- ++(.25, .3);
\draw (9.25,4.85) arc (0:90:.3);
\draw (9.45,4.85) arc (0:90:.5);
\draw (9.65,4.85) arc (0:90:.7);

\end{tikzpicture}
\label{subfig:etcma-tx}}
\hfill
\subfigure[][Receiver. Here, $\underline{\sigma}_k(l)$ is a shorthand notation for vector $\{\sigma_{k,n}^{(t)}(l)\}_{n=0}^{|\chi_{\rm TCM}|-1}$. A similar notation is used for $\tau$, $\nu$ and $\omega$.]{
\begin{tikzpicture}[scale=.7]
\draw[rounded corners=5pt, line width=2.pt] (-1.,-.5) rectangle (10.,6.5);
\node (tx) at (4.75,6.) {\bf\textsf{ETCMA Receiver}};

\draw[line width=1pt]  (6.75,3.875) rectangle (9.5,5.125);
\node at (8.125,4.5) {\small ${\rm ETCMD}_0$};
\draw[->,line width=1pt] (1.5,4.875) -- ++(2.,0) node [very near end, above left=-2pt]
{\small $\underline{\sigma}_0(l) $};
\draw[line width=1pt]  (3.5,4.5) rectangle (4.75,5.25);
\node at (4.125,4.875) {\tiny\textsf{$\rm \Pi_0^{-1}$}};
\draw[->,line width=1pt] (4.75,4.875) -- ++(2.,0.) node [very near start, above right=-2pt]
{\small $\underline{\tau}_0(l)$};
\draw[line width=1pt]  (3.5,3.75) rectangle (4.75,4.5);
\node at (4.125,4.125) {\tiny\textsf{$\rm \Pi_0$}};
\draw[<-,line width=1pt] (4.75,4.125) -- ++(2.,0.) node [very near start, above right=-2pt]
{\small $\underline{\nu}_0(l)$};
\draw[<-,line width=1pt] (1.5,4.175) -- ++(2.,0)  node [very near end, above left=-2pt]
{\small $\underline{\omega}_0(l)$};

\draw[line width=.75pt, dotted] (8.125,3.65) -- ++(.0,-.3);
\draw[line width=.75pt, dotted] (4.125,3.65) -- ++(.0,-.3);

\draw[line width=1pt]  (6.75,1.875) rectangle (9.5,3.125);
\node at (8.125,2.5) {\small ${\rm ETCMD}_k$};
\draw[->,line width=1pt] (1.5,2.875) -- ++(2.,0) node [very near end, above left=-2pt]
{\small $\underline{\sigma}_k(l)$};
\draw[line width=1pt]  (3.5,2.5) rectangle (4.75,3.25);
\node at (4.125,2.875) {\tiny\textsf{$\rm \Pi_k^{-1}$}};
\draw[->,line width=1pt] (4.75,2.875) -- ++(2.,0.) node [very near start, above right=-2pt]
{\small $\underline{\tau}_k(l)$};
\draw[line width=1pt]  (3.5,1.75) rectangle (4.75,2.5);
\node at (4.125,2.125) {\tiny\textsf{$\rm \Pi_k$}};
\draw[<-,line width=1pt] (4.75,2.125) -- ++(2.,0.) node [very near start, above right=-2pt]
{\small $\underline{\nu}_k(l)$};
\draw[<-,line width=1pt] (1.5,2.175) -- ++(2.,0)  node [very near end, above left=-2pt]
{\small $\underline{\omega}_k(l)$};
\draw[->,line width=1pt] (9.5,2.5) -- ++(1.25,0) node [very near end, above] {\small $\hat u_k(l)$};

\draw[line width=.75pt, dotted] (8.125,1.65) -- ++(.0,-.3);
\draw[line width=.75pt, dotted] (4.125,1.65) -- ++(.0,-.3);

\draw[line width=1pt]  (6.75,-.125) rectangle (9.5,1.125);
\node at (8.125,.5) {\footnotesize ${\rm ETCMD}_{K-1}$};
\draw[->,line width=1pt] (1.5,.875) -- ++(2.,0) node [very near end, above left=-3pt]
{\small $\underline{\sigma}_{K-1}(l)$};
\draw[line width=1pt]  (3.5,.5) rectangle (4.75,1.25);
\node at (4.125,.875) {\tiny $\rm \Pi_{K-1}^{-1}$};
\draw[->,line width=1pt] (4.75,.875) -- ++(2.,0.) node [very near start, above right=-2pt]
{\small $\underline{\tau}_{K-1}(l)$};
\draw[line width=1pt]  (3.5,-.25) rectangle (4.75,.5);
\node at (4.125,.125) {\tiny $\rm \Pi_{K-1}$};
\draw[<-,line width=1pt] (4.75,.125) -- ++(2.,0.) node [very near start, above right=-2pt]
{\small $\underline{\nu}_{K-1}(l)$};
\draw[<-,line width=1pt] (1.5,.175) -- ++(2.,0)  node [very near end, above left=-3pt]
{\small $\underline{\omega}_{K-1}(l)$};

\draw[line width=1pt]  (-.50,-.125) rectangle (1.5,5.125);
\node[text width=1.5cm, align=center] at (.5, 2.5) {\footnotesize \sf Enhanced Multi Stream Detector (EMSD)};

\draw[line width=1pt] (-.5,2.5) -- ++(-1.125,0.) -- ++(0., 2) -- ++(-.25, 0.3) -- ++(0.25, -0.3) -- ++(.25, .3);
\node[below] at (-1.5,2.5) {\small $r(l)$};

\end{tikzpicture}
\label{subfig:etcma-rx}}
\hfill
\caption{ETCMA transmission system.}
\label{fig:etcma-system}
\end{figure*}
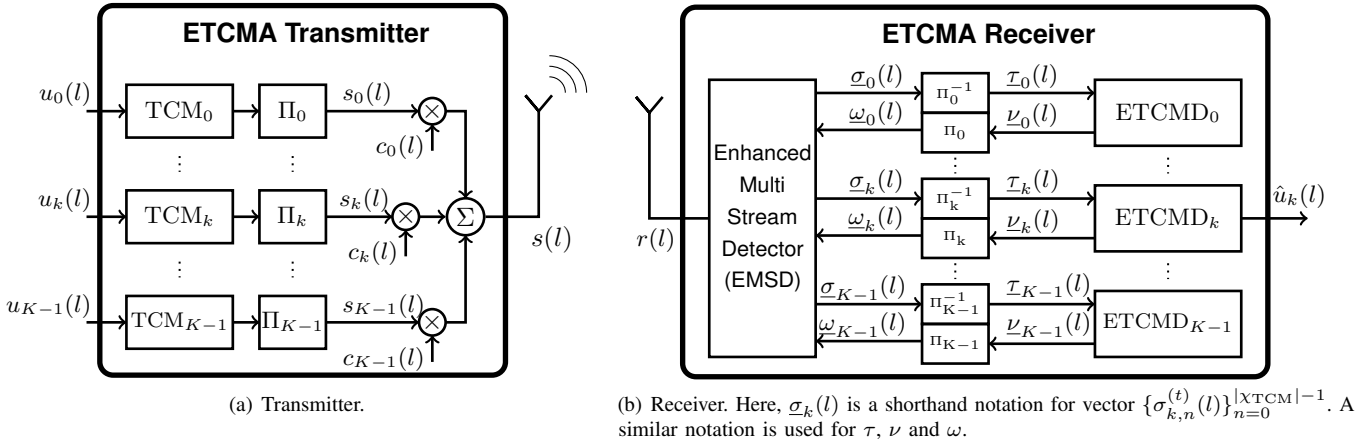

\section{The ETCMA Scheme}
\label{sec:ETCMA}

TCM has been proposed in~\cite{bib:Unger82TCM} as a coded-modulation
scheme in which the modulation is an integral part of the encoding process.
It has been shown that it is possible to achieve significant coding gains even
using simple convolutional codes with a low number of states, achieving
improved performance w.r.t. BICM~\cite{bib:BICM} when transmitting on
the AWGN channel.

In a TCMA system, each UE transmits a single TCM-encoded and
modulated data stream in the same channel resources already
used by other UEs. As a result, the received signal is the
superposition of all the transmitted signals.
In ETCMA, superposition of streams is performed at the
transmitter by a suitably designed linear combiner whose
structure is detailed hereinafter.

\subsection{ETCMA Transmitter}

The ETCMA transmitter (see Fig.~\ref{subfig:etcma-tx}) performs concurrent
transmission of multiple independent data streams, possibly intended for
different users, by a single transmitter.
Here, $K$ data streams consisting of information words of size $L$ bits are
independently encoded and modulated using
TCM encoders ${\rm TCM}_k$ then interleaved by a stream-specific
permutation $\Pi_k$. Each interleaved stream is multiplied by a specific
scrambling sequence $c_k(l)$, then multiple streams are summed and
transmitted. Stream scrambling and summing are the new features of
the ETCMA scheme w.r.t. to original TCMA.

Each stream may be encoded using different trellis codes
or modulation schemes, although we assume hereinafter that
all streams use the same TCM scheme whose parameters are
chosen as a function of $K$ using a criterion described
in Sec.~\ref{subsec:exit}.

Modulated streams are interleaved using permutations $\Pi_k$ which have
to be stream-specific in order to help the receiver to separate signals.
The set of permutations can be generated choosing $K$ distinct permutations
of the same size $L$  with uniform distribution from the set of permutations
of a certain size.
As an alternative, the permutation set can be obtained by circular shifts of a
single randomly generated permutation.
As a further alternative, Quadratic Polynomial Permutations~\cite{bib:SunTa05}
can be used.
In all cases, we observed a similar performance.
Hereinafter, we adopt random permutations.

Scrambling signatures are used to further separate streams in the
signal domain so that their discrimination is facilitated.
After interleaving, streams are scrambled and combined
before transmission.
Scrambling consists in multiplying the sequence of interleaved 
symbols $s_k(l)$  by a stream-specific signature of complex coefficients
$\mathbf{c}_k=(c_k(l))$,where $l=0,\ldots,L-1$ is the symbol index.
The transmitted signal is therefore
\begin{equation}
s(l) = \sum_{k=0}^{K-1}{c_k(l) s_k(l)}
\end{equation}
and belongs to a set of symbols
 $\chi_{\rm SUP}(l)=\{x_q(l)\}$ called
 \emph{super-constellation}\footnote{Please note that, although
 $\chi_{\rm TCM}$ does not depend on $l$, $\chi_{\rm SUP}$
may be time-varying because of the presence of different scrambling
signatures.}~\cite{bib:Brann01Globecom} whose size is
$|\chi_{\rm TCM}|^K$.
Here, $\chi_{\rm TCM}$ is the TCM constellation and $x_q(l), q = 0,
\ldots, |\chi_{\rm TCM}|^K-1$, is the $(q+1)$-th symbol of
super constellation $\chi_{\rm SUP}(l)$.

For the design of scrambling signatures, different approaches have
been undertaken.
The first approach consists in choosing the scrambling coefficients
that maximize the minimum Euclidean distance
$d_{E,\min}(\chi_{\rm SUP}(l))$ between symbols of the super-constellation.
In this case, since $\chi_{\rm TCM}$
is not time-varying, the obtained signature coefficients are independent
of time index $l$.
Their values have been found by numerical optimization.
According to a second option, we choose the coefficients as uniformly
spaced in a given interval $[0,\pi/\mu)$, where $\mu$ is a parameter whose
value depends on the shape of $\chi_{\rm TCM}$.
The upper interval endpoint $\pi/\mu$ is the smallest phase
rotation that maps $\chi_{\rm TCM}$ into itself.
We have $\mu$=1 for BPSK, $\mu$=2 for QPSK and QAM
constellations, $\mu$=4 for 8PSK, etc.
In this case, the scrambling coefficients are defined as
\[
c_k(l) = \exp(j \pi k / (K \mu)), \; \; k = 0, \ldots, K -1.
\]
A third option consists in using a set of signatures selected from
the Zadoff-Chu (ZC) class~\cite{bib:ZadoffChu}, a type of complex
sequences with low cross-correlation properties:
\begin{equation}
c_k(l)=\exp(-(j\pi r_k l(l+ L {\rm mod} 2))/L)), l=0,\ldots,L-1
\end{equation}
where $\{r_k\}$ is a set of distinct stream-specific integers
relatively prime with $L$.

For each value of $K$, the receiver performance is evaluated
by using each of the above three sequence designs.
The design resulting in the best performance is selected for inclusion
in the set of transmission parameters shown in Tab.~\ref{tab:params}.

We notice that $d_{E,\min}$ maximization results in the best
performance only when the number of streams is low,
although one could expect that, using a maximum-likelihood receiver, this would be the best approach for any number of streams.
However, in the \emph{sub-optimal} SIC receiver herein employed, the EMSD introduces
cross-stream interference which depends on the number of  streams.
Such interference is only partially removed by the ETCMD decoders.
We believe that design approaches other than $d_{E,\min}$
maximization prove more efficient in mitigating such cross-stream interference
and hence result in better performance.

\subsection{ETCMA Receiver}

A new receiver that performs significantly better than the original
TCMA receiver~\cite{bib:Brann02TCom} has been devised.
As shown in Fig.~\ref{subfig:etcma-rx}, it consists of an
Enhanced Multi-Stream Detector (EMSD) followed by a
bank of single-stream Enhanced TCM Decoders
(ETCMD)\footnote{We call our TCM decoder \emph{enhanced}
in order to remark that it is different from the original
TCM decoder used in~\cite{bib:Brann02TCom}.}.
The receiver iteratively executes the EMSD and ETCMDs
according to a Successive Interference 
Cancellation (SIC) schedule~\cite{bib:Tse}
which consists of two nested iteration loops.
The outer iterations are
indexed by variable $t=0,\ldots,N_{\rm IT}-1$, where
the number of outer iterations $N_{\rm IT}$ is
a receiver parameter.
For each outer iteration $K$ inner iterations, indexed by
variable $k=0,\ldots, K-1$, are executed.

In each inner iteration, as shown in Fig.~\ref{subfig:etcma-rx}, the EMSD calculates
a \emph{vector} of \emph{a-priori} Log-Likelihood Ratios
(LLRs) $\{\sigma_{k,n}^{(t)}(l)\}_{n=0}^{|\chi_{\rm TCM}|-1}$
for the $k$th stream modulation symbols
whose elements are computed\footnote{See~\cite{bib:ett} and references therein for
details about probability computations in the LLR domain.}
as
\begin{eqnarray}
\label{eq:sigma}
\sigma_{k,n}^{(t)}(l) & = &
 \log\dfrac{P(s_k(l) = m_n|\mathbf{r})}{P(s_k(l) = m_0|\mathbf{r})}  \nonumber \\
& = & \maxs_{q:s_k(x_q(l)) = m_n} \Lambda_{k-1,q}^{(t)}(l) \nonumber \\
 & - & \maxs_{q:s_k(x_q(l)) = m_0} \Lambda_{k-1,q}^{(t)}(l) - \omega_{k,n}^{(t-1)}(l)
\end{eqnarray}
where $m_n$ is the $n$th symbol of $\chi_{\rm TCM}$
($n=0,\ldots,|\chi_{\rm TCM}|-1$),
$x_q(l)$ is the $q$th symbol of $\chi_{\rm SUP}(l)$
($q=0,\ldots,|\chi_{\rm SUP}|-1$) and
$\mathbf{r} = (r(0), \ldots, r(L-1))$ is the received vector.
$\Lambda_{k-1,q}^{(t)}(l)$ (hereinafter called
\emph{joint LLR distribution}) represents
the LLR of the $q$th symbol
of $\chi_{\rm SUP}(l)$ updated using the \emph{a-posteriori}
extrinsic LLR of
the $n$-th TCM symbol $\omega_{k-1,n}^{(t)}$.
Here and below we use the shorthand notation
$\maxs(a, b)$ to denote $\log(e^a+e^b)$.

LLRs $\sigma_{k,n}^{(t)}(l)$ are deinterleaved obtaining
$\tau_{k,n}^{(t)}(l) = \sigma_{k,n}^{(t)}(\Pi_k^{-1}(l))$ and sent to
${\rm ETCMD}_k$ which executes the BCJR
algorithm~\cite{bib:BCJR} on the trellis of the TCM encoder
working at symbol level as in~\cite{bib:Scana01GC}
and computes updated \emph{a-posteriori} extrinsic LLRs
$\nu_{k,n}^{(t)}(l)$. After interleaving, we obtain
$\omega_{k,n}^{(t)}(l) = \nu_{k,n}^{(t)}(\Pi_k(l))$.

Finally, the EMSD updates the joint LLR distribution as
\begin{equation}
\label{eq:jointLLR}
\Lambda_{k,q}^{(t)}(l) = \Lambda_{k-1,q}^{(t)}(l) +
\omega_{k,n}^{(t)}(l)
- \omega_{k,n}^{(t-1)}(l).
\end{equation}
Here and in (\ref{eq:sigma}), when $k=0$ we set
$\Lambda_{-1,q}^{(t)}(l) =  \Lambda_{K-1,q}^{(t-1)}(l)$ for
all $q$, $l$ and $t\geq 1$.
When $k=0$ and $t=0$, we set
\begin{eqnarray}
\label{eq:jointLLRdef}
\Lambda_{-1,q}^{(0)}(l) & = & \log\dfrac{P(s(l) = x_q(l)|\mathbf{r})}{P(s(l) = x_0(l)|\mathbf{r})} \nonumber \\
& = & \dfrac{\|r(l) - x_0(l)\|^2 - \|r(l) - x_q(l)\|^2}{2 \sigma_w^2} \nonumber
\end{eqnarray}
where $x_q(l)\in \chi_{\rm SUP}(l)$ and $\sigma_w^2$ is the
variance of noise.
The LLRs $\omega_{k,n}^{(-1)}(l)$ are
set to zero for all $k$, $n$ and $l$.

After a fixed number of iterations $N_{\rm IT}$,
the receiver computes the
\emph{a-posteriori} LLRs of information bits $u_k(l)$ and
delivers the decoded information $\hat u_k(l)$ to the recipient.

\begin{figure}
\centering
\includegraphics[width=.85\columnwidth, clip=true, trim=5.2cm 1.7cm 8.4cm 6.4cm]
{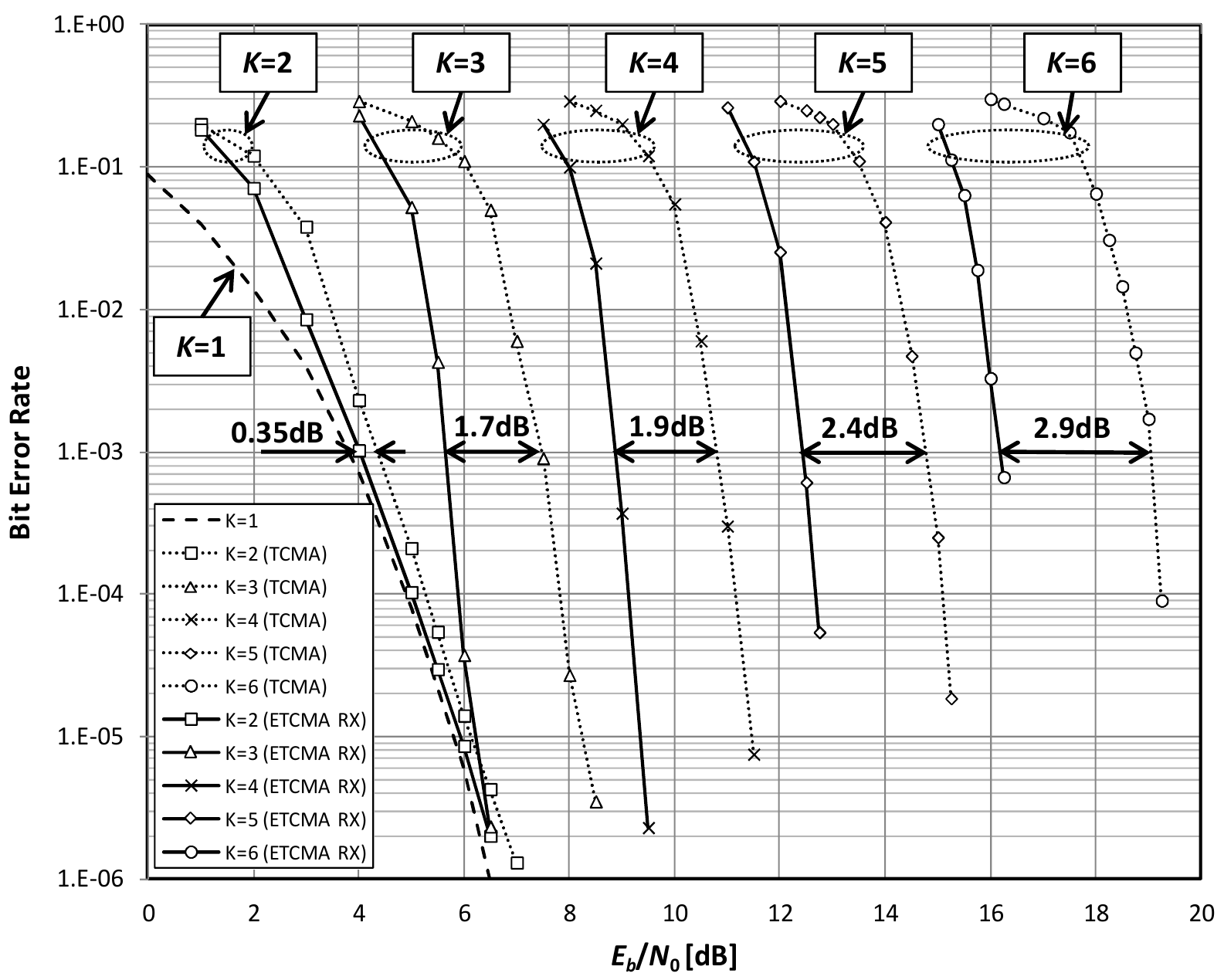}
\caption{Bit error rate of TCMA with enhanced receiver (solid curves) and
original TCMA receiver~\cite{bib:Brann02TCom} (dotted curves).
Block length is $L=1000$ bits and the receiver performs
$N_{\rm IT}=5$ iterations.}
\label{fig:etcma-ber}
\end{figure}

In the original TCMA receiver~\cite{bib:Brann02TCom}, the soft information
values exchanged through the interleavers during the iterative 
process are LLRs of \emph{coded} bits
$\beta_k^{(i)}(l) = \log[{P(d_k^{(i)}(l) = 1|\mathbf{r})}/{P(d_k^{(i)}(l) = 0|\mathbf{r})}]$.
Computing the LLRs of TCM symbols -- as in the ETCMA receiver -- instead
of coded bits  -- as in the original TCMA receiver -- results in a
slightly increased computational complexity and a significantly improved performance as shown
in Fig.~\ref{fig:etcma-ber}, where the Bit Error Rate (BER) of the ETCMA receiver
is compared with the results of~\cite{bib:Brann02TCom}.
We obtain SNR gains that increase with the number of
streams and reach 2.9 dB when $K=6$.

The overall receiver complexity is largely dominated by equations 
(\ref{eq:sigma}) and (\ref{eq:jointLLR}),
whose number of operations grows exponentially with $K$ as the size of 
the super-constellation $|\chi_{\rm SUP}| \sim |\chi_{\rm TCM}|^K$.
Simplified EMSD algorithms that compute (\ref{eq:sigma}) and (\ref{eq:jointLLR})
on suitably chosen subsets of $\chi_{\rm SUP}$
yielding significant complexity reductions are being investigated.

\subsection{TCM Encoder Design}
\label{subsec:exit}

\begin{figure}[b]
\centering
\begin{tikzpicture}[scale=.85]

\draw[line width=2.pt] (0.0,0.0) rectangle +(4.75,3.5);
\node[anchor=north] at (2.5,3.5) {${\rm TCM}_k$};

\draw[->,line width=1pt] (-.875,1.5) -- +(1.625,0.) node [near start, above=-2pt] {$u_0(l)$};
\draw[line width=1pt]  (.75,1.125) rectangle ++(.75,.75);
\node at (1.125,1.5) {\small$D$};
\draw[->,line width=1pt] (.375,1.5) -- ++(0.,0.75) -- ++(3., 0) node[near end, above]
{\small $d_k^{(1)}(l)$};
\draw[->,line width=1pt] (.375,1.5) -- ++(0.,-0.75) -- ++(1.375, 0);
\draw[->,line width=1pt] (1.5,1.5) -- ++(0.5, 0) -- ++(0.,-0.5);
\draw[line width=1pt] (2.,.75) circle (.25);
\draw[line width=1pt] (1.75,.75) -- +(.5,0.);
\draw[line width=1pt] (2.,.5) -- +(0.,.5);
\draw[->,line width=1pt] (2.25,.75) -- +(1.125,0.)  node[very near end, above left]
{\small $d_k^{(2)}(l)$};

\draw[line width=1pt]  (3.375,.5) rectangle ++(1.,2.);
\node at (3.875,1.5) {\footnotesize \sf QPSK};

\draw[->,line width=1pt] (4.375,1.5) -- +(1.,0.)  node[very near end, above]
{\small $s_k(\tilde l)$};

\end{tikzpicture}
\caption{TCM encoder-modulator. The two-state trellis encoder
has generator coefficients $(2,3)_8$.
Here, $\tilde l = \Pi_k^{-1}(l)$.}
\label{fig:tcm-encoder}
\end{figure}
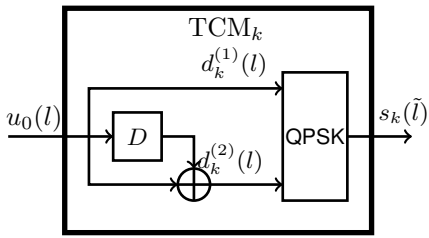

\begin{figure}[t]
\centering
\includegraphics[width=.75\columnwidth,clip=true,trim=2.1cm 8.1cm 2.3cm 8.3cm]
{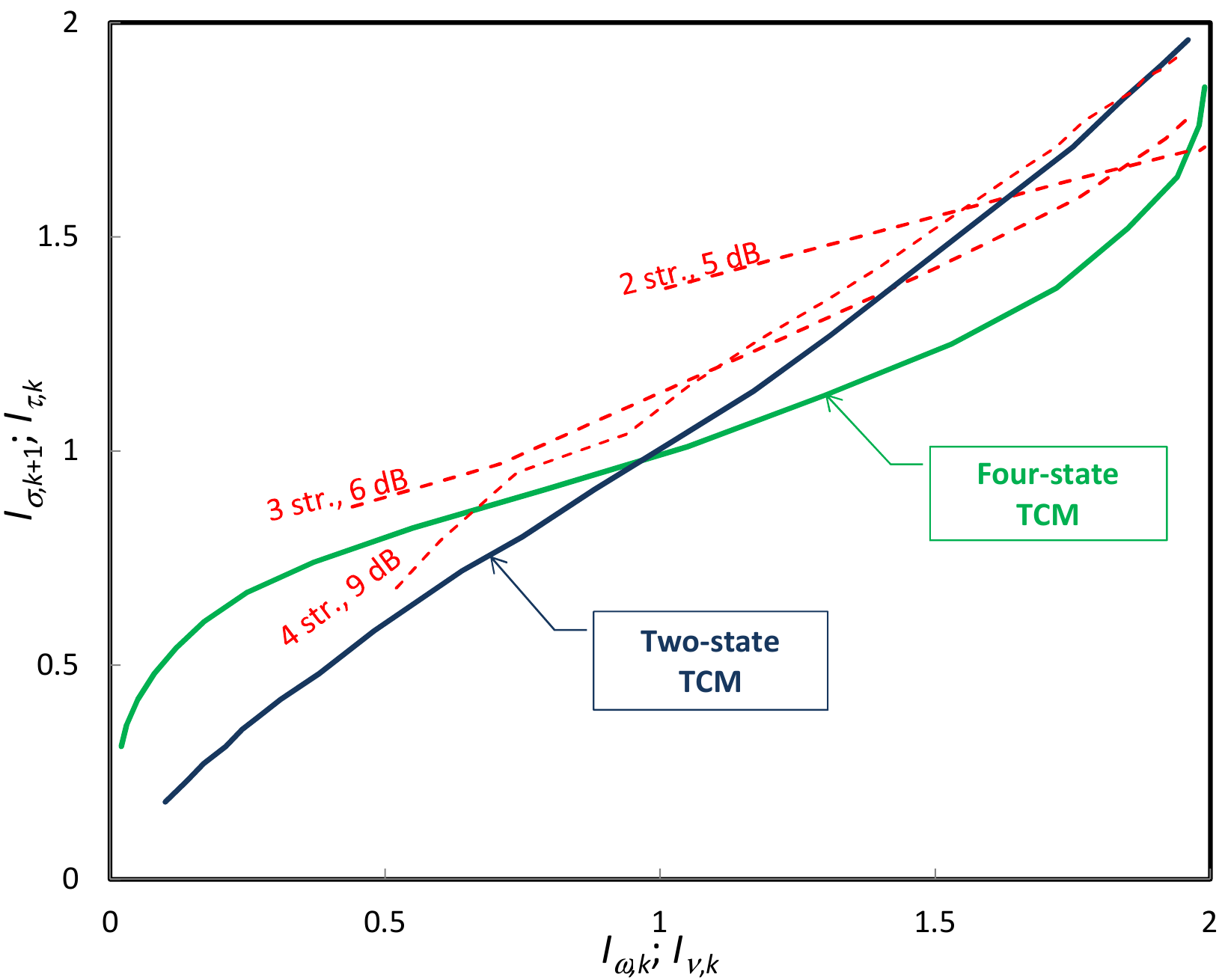}
\caption{EXIT charts of ETCMA receiver. Solid curves refer to the ETCMDs;
dashed curves refer to the EMSD.}
\label{fig:exit}
\end{figure}

The employed TCM scheme depends on the overloading factor $K$ and consist of a rate 1/2 convolutional encoder (CE)
connected to a QPSK symbol mapper.
We consider two CE schemes: a four-state CE with generator coefficients
$(5,7)_8$ and a two-state CE with generator coefficients $(2,3)_8$
(see Fig.~\ref{fig:tcm-encoder}).

Using EXtrinsic Information Transfer (EXIT)
charts~\cite{bib:tenBr01TranCom, bib:Brann02GC}, the behavior of the
ETCMA receiver has been analyzed, yielding the conclusion that, when
$K \leq 3$ four state encoders result in better performance whilst
for $K \geq 4$ two-state encoders should be used.
The EXIT charts, described in terms of average mutual information
functions $I_{\sigma,k}^{(t)}$, $I_{\tau,k}^{(t)}$, $I_{\nu,k}^{(t)}$ and
$I_{\omega,k}^{(t)}$, are shown in Fig.~\ref{fig:exit}.
Solid curves represent the ETCMD characteristics ($I_{\tau,k}$ vs. $I_{\nu,k}$),
while dashed curves represent the characteristics of the EMSD
($I_{\sigma, k+1}$ vs. $I_{\tau,k}$), which depend both on the SNR and on $K$.
Using an approach similar to~\cite{bib:Scana01GC}, the average mutual information
$I_{\omega,k}^{(t)}=\mathbb{E}_{l}[I(s_k(l);\mathbf{\omega}_k^{(t)}(l))]$
between the modulation symbols $s_k(l)$ and the LLRs
$\mathbf{\omega}_{k,n}^{(t)}(l)$
at iteration $t$ has been estimated as
\begin{equation*}
I_{\omega,k}^{(t)} = \log_2 |\chi_{\rm TCM}| - \frac{1}{L}\sum_{l=0}^{L-1}
{(\maxs_n (\omega_{k,n}^{(t)}(l)) - \omega_{k,\bar n_{k,l}}^{(t)}(l))}
\end{equation*}
where $\bar n_{k,l}$ is the index of the transmitted symbol:
\mbox{$s_k(l) = m_{\bar n_{k,l}}$}.
Similarly, we estimated $I_{\sigma,k}^{(t)}$, $I_{\tau,k}^{(t)}$
and $I_{\nu,k}^{(t)}$.

If an EMSD curve and an ETCMD curve intersect each other in a point close to
the upper right corner -- where $I_\sigma = I_\omega = 2$ -- then the decoder will
converge to the correct code word and deliver an error-free information word.
This is the case of the EMSD characteristic for $K=2$ streams at SNR=5 dB
(upper red dashed curve of Fig.~\ref{fig:exit}) and the four-state ETCMA
characteristic (solid green curve in Fig.~\ref{fig:exit}).
Choosing the two-state TCM (solid blue curve in Fig.~\ref{fig:exit}), the intersection
would be moved to a point with lower $I_\sigma$ and $ I_\omega$, thus predicting
a higher error rate.
The same considerations lead to the choice of the four-state encoder for $K=3$.
When $K=4$, at SNR of 9 dB the EMSD characteristics intersects the four-state
ETCMD characteristic at rather low values $I_\sigma$ and $ I_\omega$, whilst
the intersection with the two-state ETCMD characteristic is much closer to 
$I_\sigma = I_\omega = 2$.  A similar behavior is also expected when $K \geq 5$.
As a result, when $K \geq 4$ we choose the two-state encoder.

\section{Results}
\label{sec:results}

Simulations have been performed in order to assess the performance
of ETCMA schemes.
The considered channel model is AWGN with two-sided power
spectral density of noise $G_w(f) = N_0/2$.
The BLock Error Rate (BLER) has been evaluated and the aggregate SE
has been computed as
\begin{equation}
\label{eq:se}
SE(K)=(1-{ \rm BLER})Rm_0 K \; \; \rm [bits/s/Hz].
\end{equation}
Here,  $R=1/2$
is the channel code rate and $m_0 = 2$ is the modulation order. 
The asymptotic aggregate SE (ASE) is
\begin{equation}
SE_\infty(K)=\lim_{\rm SNR\to\infty}SE(K).
\end{equation}
The most relevant metric we take into account is the
single-stream SNR loss $\Delta_{\rm SNR}(K,\rho)$, which is the increase of
SNR with respect to the single-stream SNR required to
achieve a given ratio $\rho$ of the ASE when the overloading factor is $K>1$:
\begin{eqnarray}
\label{eq:snr-loss}
\Delta_{\rm SNR}(K,\rho) & = & \left.{\rm SNR}\right|_{SE(K)=\rho SE_\infty (K)}  \nonumber \\
 & - &\left.{\rm SNR}\right|_{SE(1)=\rho SE_\infty (1)}.
\end{eqnarray}

Each information word is transmitted using $L=240$
REs.
The transmission parameters are selected according to
Tab.~\ref{tab:params}.

\begin{table}
\centering
\caption{Transmission parameters.}
\begin{tabular}{cccc}
\multirow{2}{*}{$K$} & \bfseries TCM constraint & \bfseries TCM & \bfseries Scrambling \\
& \bfseries length & \bfseries modulation & \bfseries sequence type\\
\midrule
2	& 3 & QPSK & Max. $d_{E,min}$ \\
3	& 3 & QPSK & Uniformly spaced \\
4, 5	& 2 & QPSK & Uniformly spaced \\
6, 7	& 2 & QPSK & Zadoff-Chu \\
\end{tabular}
\label{tab:params}.
\end{table}


The receiver performs $N_{\rm IT}$ outer iterations.
Decoding complexity has been taken into account for the
determination of the number of iterations.
When the four-state TCM encoder is used, the receiver executes
$N_{\rm IT} = 10$ iterations.
Using two-state encoders, the trellis complexity of TCM is reduced,
therefore we increase the number of iterations to $N_{\rm IT} = 15$ in
order to partially compensate for the reduced complexity of ETCMDs.
\begin{figure*}[t]
\centering
\hfill
\subfigure[][Aggregate SE of ETCMA (solid curves) compared with the
SE obtained using the original TCMA receiver~\cite{bib:Brann02TCom} (dashed curves).]{
\includegraphics[scale=.35,clip=true,trim=4.2cm 2.7cm 4.cm 2.8cm]{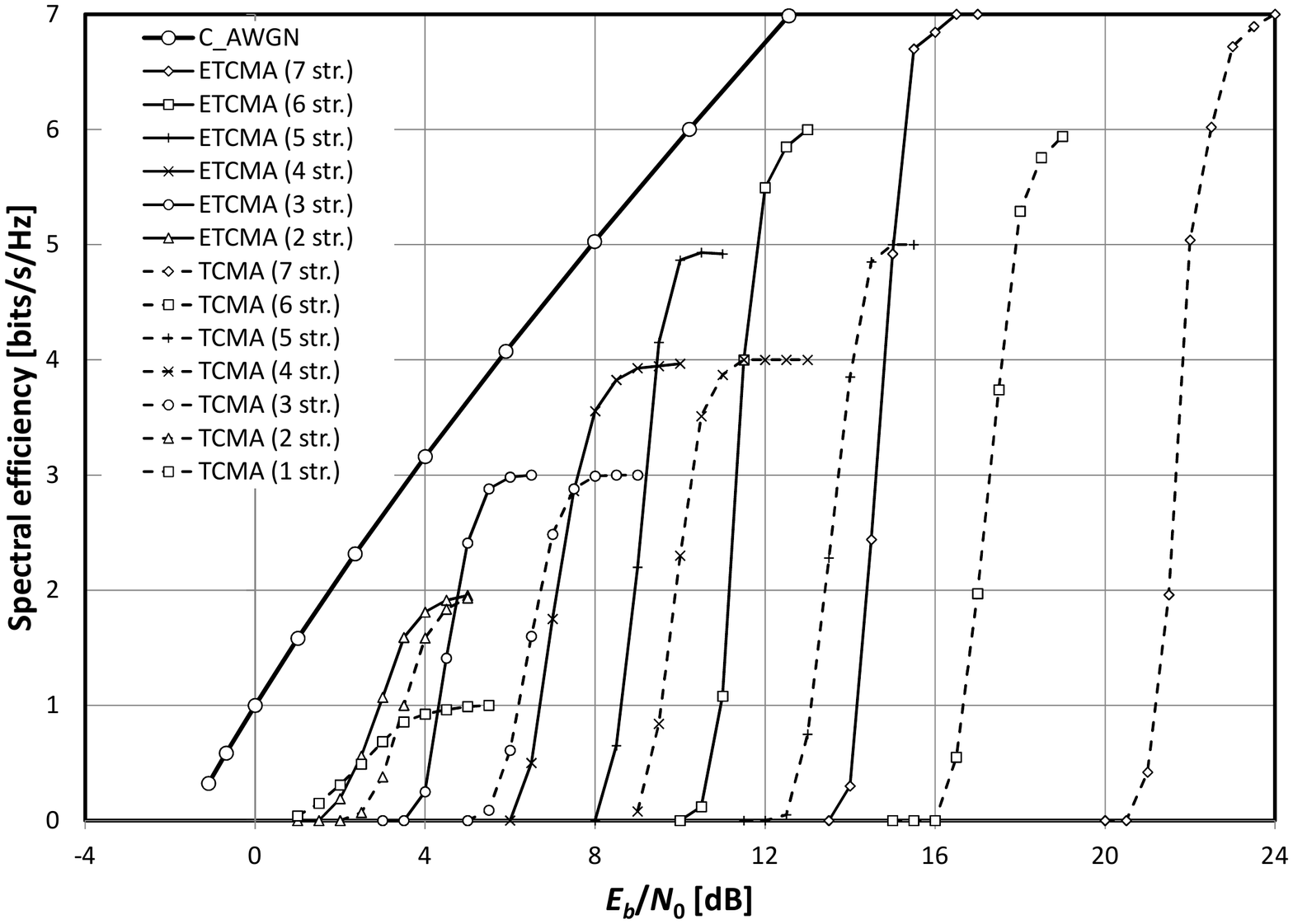}\label{subfig:etcma-se}}
\hfill
\subfigure[][SE of ETCMA (solid curves) compared with turbo coded LTE
(dashed curves).]{
\includegraphics[scale=.35,clip=true,trim=3.8cm 2.6cm 4.1cm 2.9cm]{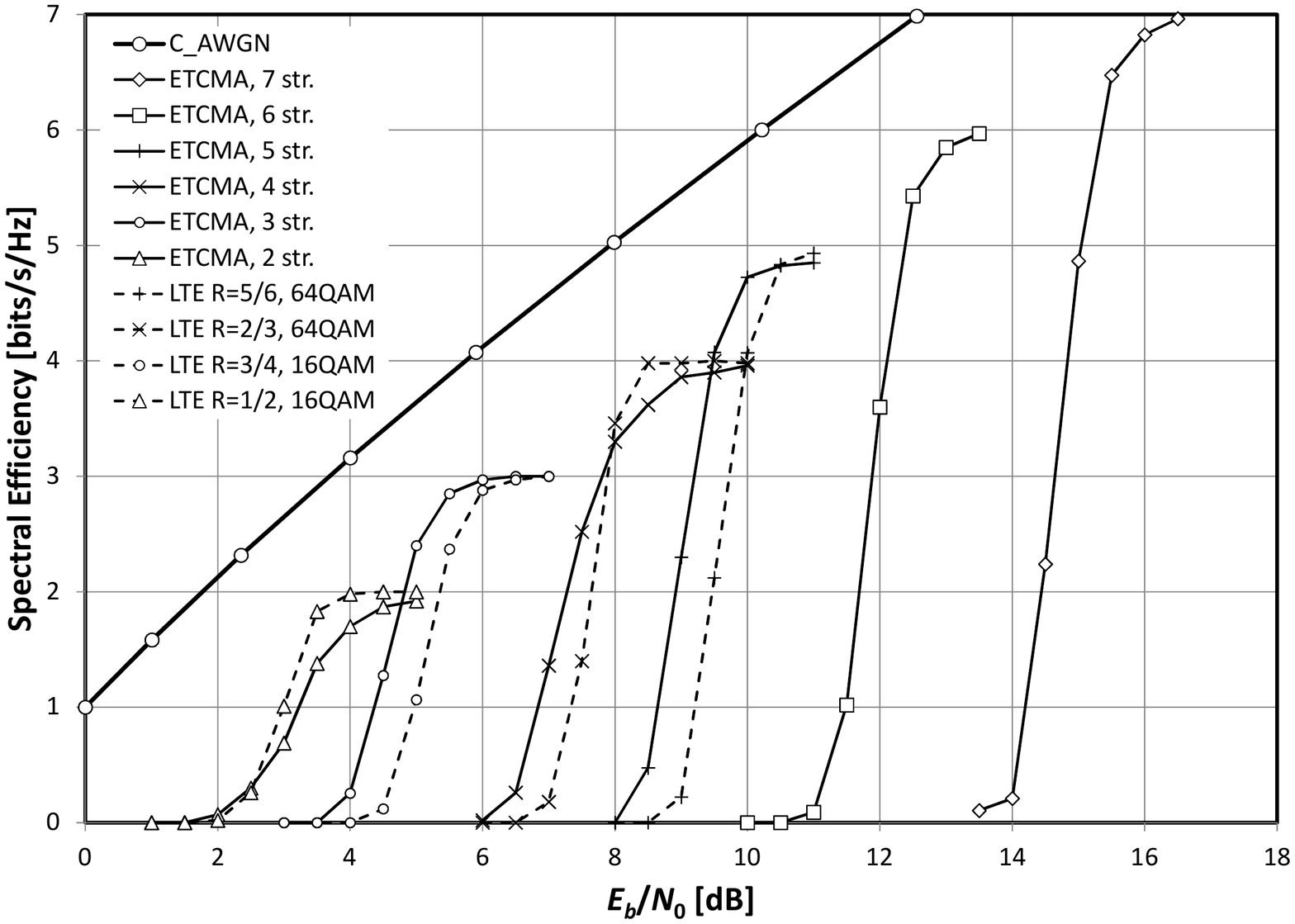}
\label{subfig:etcma-se-lte}}
\hfill
\caption{Spectral efficiency of ETCMA.}
\label{fig:etcma-performance}
\end{figure*}
Fig.~\ref{fig:etcma-performance} shows the SE of ETCMA on the AWGN channel. 
In Fig.~\ref{subfig:etcma-se}, the SE of ETCMA is compared with the SE
of the original TCMA~\cite{bib:Brann02TCom}.
The SNR loss of ETCMA computed applying (\ref{eq:snr-loss}) is
summarized in Tab.~\ref{tab:snr-loss} and therein compared with the
SNR loss fo TCMA.
The observed SNR gains\footnote{The SNR gains here reported are different from
those shown in Fig.~\ref{fig:etcma-ber} because of the different number of
receiver iterations.}
increase with $K$ and range from less than
0.35 dB for $K=2$ to a remarkable 7.15 dB for $K=7$.
The AWGN channel capacity is also shown
as a reference.
We observe that ETCMA exhibits a much smaller SNR gap
of about $2\div2.5$ dB w.r.t. the
AWGN capacity.

Fig.~\ref{subfig:etcma-se-lte} compares the SE of ETCMA with the SE of a
turbo coded LTE link~\cite{bib:LTEr12-212}.
Both systems use $L=240$ REs per
data block.
In the LTE system, an information word of $K L$ bits is
transmitted using $L$ REs.
The ETCMA transmitter segments the information word
into $K$ sub-words of length $L$ which are independently encoded and 
combined then transmitted.
At the receiver, the inverse procedure is performed in order to
reassemble the information word.
SE is evaluated using (\ref{eq:se}) but here, as a difference w.r.t.
Fig.~\ref{subfig:etcma-se},  BLER is computed over blocks of size $KL$ bits.
ETCMA exhibits a higher SE than LTE for several SNR values and
achieves SE of 6 bits/s/Hz and above that are
not achievable with current LTE systems, while performing 
close to the AWGN channel capacity.

\begin{table}
\centering
\caption{Single-stream SNR losses $\Delta_{\rm SNR}(K,\rho)$  of TCMA
and ETCMA for $\rho=0.9$. Rightmost column: SNR gain of ETCMA over TCMA.}
\begin{tabular}{cccc}
$K$ & \bfseries TCMA [dB] & \bfseries ETCMA [dB] & \bfseries SNR gain [dB]\\
\midrule
2	& 0.6 & 0.25 & 0.35 \\
3	& 3.4 & 1.55 & 1.85 \\
4	& 6.75 & 4.3 & 2.45 \\
5	& 10.45 & 6.0 & 4.45 \\
6	& 14.25 & 8.2 & 6.05 \\
7	& 18.8 & 11.65 & 7.15
\end{tabular}
\label{tab:snr-loss}
\end{table}

\section{Conclusions}
\label{sec:conclusions}

We have proposed a new OLMA 
scheme that, compared to original TCMA, provides
large SNR gains of up to 7.15 dB at 7 bits/s/Hz.
In terms of aggregate SE, it approaches the capacity of the AWGN
channel within a SNR gap of about 2 dB over a wide range of
spectral efficiencies.


\end{document}